\documentclass[12pt]{iopart}
\usepackage{xcolor}

\usepackage{url}
\usepackage[super]{cite}
\usepackage{appendix}
\usepackage{amsfonts}
\begin{document}
\title{Analytic corrections to  AdS by scalar matter and  curvature squared term}
\author{Lata Kh Joshi $^*$}
\ead{latamj@phy.iitb.ac.in}
\author{P. Ramadevi $^*$}
\ead{ramadevi@phy.iitb.ac.in}

\address{$^*$Indian Institute of Technology Bombay, Powai, Mumbai, Maharashtra, India 400076}
\maketitle
\begin{abstract}
We revisit the background solution for scalar matter coupled higher derivative gravity originally reported in arXiv: 1409.8019[hep-th]. 
In this letter, we choose a convenient ansatz for metric which determines the first order perturbative corrections to scalar as well as geometry.

\end{abstract}
\maketitle
AdS-CFT correspondence \cite{Maldacena, Witten}  has been enormously successful in predicting transport properties of strongly coupled systems. One of the most celebrated result obtained using this tool is the ratio of shear viscosity ($\eta$) to entropy density ($s$), stating: $\eta/s \geq 1/4 \pi$ \cite{Policastro:2001yc, Kovtun:2004de}.  
The saturation and violation of the  bound \cite{Buchel:2003tz,Brigante:2007nu, kp, Banerjee:2009fm, Cai:2009zv, Cremonini:2011ej, myers} depends on the gravity theories 
having two-derivative  and higher derivative curvature terms respectively. These higher derivative curvature terms naturally arise when we include stringy corrections ($\alpha'$).

The temperature behaviour of $\eta/s$, which depends 
on the  scalar field and charged matter field couplings to higher derivative curvature terms, was studied in Refs.\cite{Cremonini:2011ej, sera}. In particular, they obtained corrections 
to $\eta/s$ due to four-derivative gravity terms using the background for 
matter coupled Einstein action. For theoretical completeness,  the geometry 
and matter field must be corrected (backreaction) due to higher derivative curvature terms. 
Similar to the backreaction works for higher derivative gravity\cite{kp} and charged matter coupled higher derivatives \cite{myers}, we attempted the backreaction solution for scalar matter coupled higher derivative gravity\cite{joshi}.

The choice of ansatz in Ref. \cite{joshi} was motivated by the solutions in Refs. \cite{kp,lee}. Though, we could find the backreaction of metric, we could not  determine correction to scalar matter. Our main aim in this letter is  to deduce both metric and matter corrections by taking a suitable ansatz. 

We briefly recapitulate the essential data of our earlier work \cite{joshi}. The scalar matter coupled 
higher derivative gravity action, in $D$ dimensional spacetime, under investigation is 
\begin{equation}
\label{action}
\hskip-1cm S = \frac{1}{2\kappa_D^2} \int{d^{D}x \sqrt{-g} \left(R-\frac{4}{D-2}(\partial \Phi)^2-V(\Phi) + L^2 \beta G(\Phi)R_{\mu \nu \sigma \rho}R^{\mu \nu \sigma \rho}\right)}
\end{equation}
where we take, $V(\Phi)= 2 \Lambda e^{\alpha \Phi(r)}$, $\Lambda < 0$ and $G(\Phi)= e^{\gamma \Phi(r)}$. Note that the coupling $\beta$, in the higher derivative gravity term, must be 
taken small ($|\beta|<1$) so as to obtain perturbative corrections to the Einstein matter.   We take  AdS radius $L$ as unity and $\alpha=-\gamma$ to find the ${\mathcal O}(\beta)$ backreaction effects on geometry and the scalar field. Formally, the metric representing the geometry 
can be expanded as $g_{\mu \nu}=g_{\mu \nu}^{(0)}+\beta g_{\mu \nu}^{(1)}$ and the scalar field as $\Phi=\Phi^{(0)}+\beta \Phi^{(1)}$. Note that the superscript keeps track of order of
$\beta$  in the expansion. 

The zeroth order ($\mathcal O(\beta^0)$) parts, $g_{\mu \nu}^{(0)}$ and $\Phi^{(0)}$, are the solutions for matter coupled Einstein action\cite{lee, joshi}.  In this work, we will determine $g_{\mu \nu}^{(1)}$ and $\Phi^{(1)}$  to have full analytic solution upto first order in  $\beta$.

The solution which we obtain must converge with the known results in suitable limits namely, (i) $\Phi=0,\beta=0$ (Schwarzchild AdS) (ii) $\Phi=0$\cite{kp} and (iii) $\beta=0$\cite{lee}.

We modify the ansatz in Ref. \cite{joshi} by introducing $c_{\pm}(r,\beta)$ as follows:
\begin{equation}
\label{ansatz}
 ds^2 = -r^{-2a}(1-r^{c_+(r,\beta)})dt^2+\frac{dr^2}{r^{-2a}(1-r^{c_-(r,\beta)})r^4}+r^{-2a}d\vec{x}^2~;~
\Phi(r,\beta)~,
\end{equation}
with $c_\pm$ and $\Phi$ as,
\begin{equation}
c_\pm(r,\beta) = c_0+\frac{\log[1-\beta (\kappa(r)\pm\delta(r))]}{\log(r)}~;~\Phi(r,\beta) = m \log(r)+\beta \xi (r)~,
\end{equation}\label{back}
and,
\begin{eqnarray}
\label{acm}
a &=&\frac{16}{P}~;~
\Lambda =\frac{8 (D-2) Q}{P^2}~;~
m=\frac{-2 S }{\gamma P}~;~
c_0 = -\frac{Q}{P}\label{Lamb}~,
\end{eqnarray}
where,
\begin{eqnarray}
 S=(D-2)^2~\gamma^2~,~P=16+(D-2)^2\gamma^2~ \mathrm{and}~Q=S-16(D-1)~.
\end{eqnarray}
Observe that the ansatz still preserves $SO(D-2)$ symmetry in the $\{\vec{x}\}$ coordinates,The solution in the absence of higher derivatives, i.e. the term $g_{\mu \nu}^{(0)}$, has the horizon at $r=r_h=1$. Naively, one expects a shift in horizon such that $r_h=1+\beta r_1$ when higher derivative terms are taken into account\cite{kp}. This shift in the horizon is determined by looking at the zero (pole) in $g_{tt}$ $(g_{rr})$ component of the metric.
Alternatively, one can fix the horizon at $r_h=1$. The difference in the location of
horizon will only  shift the metric components by a constant.

By varying action (\ref{action}) with respect to metric field $g^{\mu \nu}$ and scalar field $\Phi(r)$, we have a set of four independent Euler-Lagrange equations \footnote{See appendix A in Ref. \cite{joshi} for the set of equations.}.
To determine $\delta(r), \kappa(r)$ and $\xi(r)$, we focus on the first order in $\beta$
terms in these  coupled differential equations. We take a suitable linear combination of these coupled equations so that one differential equation involves only $\xi(r)$. This equation has the form:
\begin{eqnarray}
\Box^{(0)} \xi = \frac{1}{\sqrt{-g^{(0)}}}\partial_r\left(\sqrt{-g^{(0)}} g^{(0) rr} \partial_r \xi(r)\right) &=& f(r)\label {xidef}\\
~~~~~~~~~~~~~~~~~~~~~~~~\Rightarrow  \sqrt{-g^{(0)}} g^{(0) rr} \xi^\prime(r)  &=& \int{dr\sqrt{-g^{(0)}} f(r) }~.\label{xiprimeeq}
\end{eqnarray}
Note that  $f(r)$ contains all the terms appearing from potential-like terms and from the metric.  
The presence of the metric component $g^{(0) rr}$, which vanishes at $r=1$, in the LHS of eqn (\ref{xiprimeeq}) forces the RHS to be zero at $r=1$. Using this, we fix the first  integration constant for the
second order $\xi$ equation (\ref{xidef}) \footnote{We thank Sayantani Bhattacharyya for clarifying this crucial step to us.}.  Next, integrating eqn (\ref{xiprimeeq}) we find the analytic form for $\xi$, where the integration constant is fixed by demanding that the total matter field goes to zero at horizon radius $r=r_h=1$.

The solution for leading order matter correction $\xi(r)$ under above boundary values is,
\begin{eqnarray}
\xi(r)=\frac{-4 S}{r \gamma Q P^3 }&\left[P \{S^2-16 S (D-4)+2^8 (D-3)\} \left(r^N-r\right)+\right.\nonumber\\
&\left.~~2^6  r  (P-24) Q \log (r)\right]~~\label{xi}
\end{eqnarray}
where,
$ N=a D=16D/P$.

Having known the form for $\xi(r)$ in eqn (\ref{xi}), we can combine remaining Einstein equations to find first order differential equation in the metric correction $\delta(r)$. 
The solution turns out to vanish at $r=1$ irrespective of the integration constant.
Comparing the solution in Ref. \cite{kp} with our metric ansatz, we require $c_+(r,\beta)=
c_-(r,\beta)$ when $\gamma=0$. Hence we fix the integration constant using this $\gamma=0$ limit. The form for $\delta(r)$ turns out to be,
\begin{eqnarray}
\delta(r)=\frac{4 S \left(1-r^{1-N}\right)}{r Q P^3}& \left[r^{N}\left(S^3-2^4 S^2 (D-3) +2^{12}(D-2)\right) +\right.\nonumber\\
&\left.~~r~ 2^5  \left(P-24\right)Q \log (r)\right]. \label{delta}
\end{eqnarray}
It is now straight forward to use $\xi(r)$ and $\delta(r)$ to find first order differential equation for $\kappa(r)$. Again, to fix the integration constant we demand the horizon to be at $r=1$. Following is the last unknown $\kappa(r)$,
\begin{eqnarray}
\kappa(r)&=&\frac{2r^{-1-N}}{P^3(D-2) Q } \left[P\left\{r^N (2-D)\left( S^3-8 S^2 (3 D-10)+2^8 (D-5) S\right.\right.\right.\nonumber\\
&&\left. \left.\left. +2^{11}(D-4) (D-3) (D-1)\right)+2 r \left(- 16 S^2 \left(D^2-9 D+15\right)\right.\right.\right.\nonumber\\
&&\left.\left.\left.+2^{11} (D-4) (D-1)+ S^3 (D-2)-2^8 S \left(D^2-D-4\right)\right)\right\}\times\right.
\nonumber\\
&&\left. \left(r^N-r\right)+2^6 S (D-2) r^2 Q (P-24) \log (r)\right]. \label{kappadelta}
\end{eqnarray}

To summarize, we attempted to see the backreaction 
to the AdS background in the presence of a simple potential $V(\Phi)= e^{\alpha \Phi}$
and higher derivative matter coupling $G(\Phi)=e^{\gamma \Phi}$.
Interestingly, we find a closed form solution (\ref{xi}, \ref{delta}, \ref{kappadelta}) for $\alpha=-\gamma$.
Ideally, we should attempt finding backreaction solution for 
QCD like potentials $V(\Phi)$ \cite{Gubser:2008yx}. This appears to be a challenging problem to do analytically. 

Phenomenolgically, scalar matter coupled higher derivative gravity actions have interesting results on shear viscosity \cite{sera}. Shear viscosity and entropy density for a field theory dual to action (\ref{action}) are reported in the appendix (D) of Ref.\cite{joshi}.  For the boundary conditions chosen in this letter where the horizon is kept intact at  $r=r_h=1$, there is no backreaction correction to  $\eta$ and $s$. If the location of horizon is $\beta$ dependent\cite{joshi}, we will get backreaction corrections to $\eta$ and $s$. An important observation at this point is that the ratio $\eta/s$ is same whether the horizon is shifted or not due to backreaction.

Mathematically, for the action (\ref{action}) we could use the backreaction solution (\ref{ansatz}, \ref{xi}, \ref{delta}, \ref{kappadelta}) and determine 
second order corrections to both $\eta$ and $s$. There are subtle issues 
at $\mathcal O(\beta^2)$. Recall the four-derivative term is due to stringy correction
$\alpha'\equiv \beta$. In order to go to next order in perturbation, we need to add a six-derivative term ($\alpha'^2$) in the action. Further, we did not include other possible four-derivative terms like $R^2$ and $R_{\mu \nu}R^{\mu \nu}$ as they can be removed by field redefinition at $\mathcal O(\beta)$ \cite{Banerjee:2009fm} . However these terms will be important at $\mathcal O(\beta^2)$. To sum up, we need to incorporate all these factors to get a complete result on $\eta/s$ at $\mathcal O(\beta^2)$.

\section*{Acknowledgement}
We would like to thank Sayantani Bhattacharyya, Suvankar Dutta and Urjit Yajnik for various helpful discussions. We thank Utkarsh Sharma for the collaborations in the initial stages of this project. LKJ would like to thank organizers of \textit{Indian Strings Meet-14, Puri, India} and \textit{Applications of AdS/CFT to QCD and CMT, CRM Montreal, Canada}, where parts of this work were discussed.  
\section*{References}
\bibliographystyle{unsrt}
\bibliography{biblio}
\end{document}